\documentclass[twocolumn]{aastex631}

\usepackage{natbib}
\usepackage{amsmath}
\usepackage{multirow}
\usepackage[normalem]{ulem}
\usepackage{xcolor}
\usepackage{xspace}
\usepackage{enumitem}   
\usepackage{graphicx}
\usepackage{wrapfig}
\usepackage[running,mathlines]{lineno}
\usepackage[super]{nth}
\usepackage{longtable}
\usepackage{tabularx}

\newcommand{\Rearth}{$R_\oplus$\xspace}
\newcommand{\Mearth}{$M_\oplus$\xspace}
\newcommand{\Rsun}{$R_\odot$\xspace}
\newcommand{\Msun}{$M_\odot$\xspace}

\newcommand{\gaia}{\texttt{Gaia}\xspace}
\newcommand{\tess}{\texttt{TESS}\xspace}

\newcommand\Mjup{M$_{\rm{Jup}}$\xspace}
\newcommand\Rjup{R$_{\rm{Jup}}$\xspace}
\newcommand\vsini{$v$~sin~$i$}

\newcommand{\unit}[1]{\ensuremath{\, \mathrm{#1}}}
\newcommand{\arcsecond}{$''$\xspace}

\newcommand{\PSUAA}{Department of Astronomy \& Astrophysics, 525 Davey Laboratory, 251 Pollock Road, Penn State, University Park, PA, 16802, USA}
\newcommand{\PSUCEHW}{Center for Exoplanets and Habitable Worlds, 525 Davey Laboratory, 251 Pollock Road, Penn State, University Park, PA, 16802, USA}

\newcommand{\PSETI}{Penn State Extraterrestrial Intelligence Center, 525 Davey Laboratory, 251 Pollock Road, Penn State, University Park, PA, 16802, USA}

\shorttitle{TOI-5573b}
\shortauthors{Fernandes et al. 2025b}

\begin{document}

\title{\textit{Searching for GEMS:} Confirmation of TOI-5573\,b, a Cool, Saturn-like Planet Orbiting An M-dwarf\footnote{Based on observations obtained with the Hobby-Eberly Telescope (HET), which is a joint project of the University of Texas at Austin, the Pennsylvania State University, Ludwig-Maximillians-Universitaet Muenchen, and Georg-August Universitaet Goettingen. The HET is named in honor of its principal benefactors, William P. Hobby and Robert E. Eberly}}

\correspondingauthor{Rachel B. Fernandes}
\email{rbf5378@psu.edu}

\author[0000-0002-3853-7327]{Rachel B. Fernandes}
\altaffiliation{President's Postdoctoral Fellow}
\affiliation{\PSUAA}
\affiliation{\PSUCEHW}

\author[0000-0001-8401-4300]{Shubham Kanodia}
\affiliation{Earth and Planets Laboratory, Carnegie Science, 5241 Broad Branch Road, NW, Washington, DC 20015, USA}

\author[0000-0003-1439-2781]{Megan Delamer}
\affiliation{\PSUAA}
\affiliation{\PSUCEHW}

\author[0009-0000-1825-4306]{Andrew Hotnisky}
\affiliation{\PSUAA}
\affiliation{\PSUCEHW}

\author[0000-0002-7127-7643]{Te Han}
\affiliation{Department of Physics \& Astronomy, The University of California, Irvine, Irvine, CA 92697, USA}

\author[0000-0003-4835-0619]{Caleb I. Cañas}
\altaffiliation{NASA Postdoctoral Fellow}
\affiliation{NASA Goddard Space Flight Center, 8800 Greenbelt Road, Greenbelt, MD 20771, USA}

\author[0000-0002-2990-7613]{Jessica Libby-Roberts}
\affiliation{\PSUAA}
\affiliation{\PSUCEHW}

\author[0009-0006-7298-619X]{Varghese Reji}
\affiliation{Department of Astronomy and Astrophysics, Tata Institute of Fundamental Research, Homi Bhabha Road, Colaba, Mumbai 400005, India}

\author[0000-0002-5463-9980]{Arvind F. Gupta}
\affiliation{U.S. National Science Foundation National Optical-Infrared Astronomy Research Laboratory, 950 N.\ Cherry Ave., Tucson, AZ 85719, USA}

\author[0000-0003-0353-9741]{Jaime A. Alvarado-Montes}
\affiliation{Australian Astronomical Optics, Macquarie University, Balaclava Road, North Ryde, NSW 2109, Australia}
\affiliation{Astrophysics and Space Technologies Research Centre, Macquarie University, Balaclava Road, North Ryde, NSW 2109, Australia}

\author[0000-0003-4384-7220]{Chad F. Bender}
\affiliation{Steward Observatory, University of Arizona, 933 N. Cherry Ave, Tucson, AZ 85721, USA}

\author[0000-0002-6096-1749]{Cullen H. Blake}
\affiliation{Department of Physics and Astronomy, University of Pennsylvania, 209 S 33rd Street, Philadelphia, PA 19104, USA}

\author[0000-0001-9662-3496]{William D. Cochran}
\affiliation{McDonald Observatory and Center for Planetary Systems Habitability, The University of Texas at Austin, Austin, TX 78730, USA}

\author[0000-0002-7564-6047]{Zo\"{e} L. de Beurs}
\affiliation{NSF Graduate Research Fellow, MIT Presidential Fellow, MIT Collamore-Rogers Fellow, MIT Teaching Development Fellow}
\affiliation{Department of Earth, Atmospheric and Planetary Sciences, Massachusetts Institute of Technology, 77 Massachusetts Avenue, Cambridge, MA 02139, USA}

\author[0000-0002-2144-0764]{Scott A. Diddams}
\affiliation{Electrical, Computer \& Energy Engineering, University of Colorado, 425 UCB, Boulder, CO 80309, USA}
\affiliation{Department of Physics, University of Colorado, 2000 Colorado Avenue, Boulder, CO 80309, USA}

\author[0000-0002-3610-6953]{Jiayin Dong}
\affiliation{Center for Computational Astrophysics, Flatiron Institute, 162 Fifth Avenue, New York, NY 10010, USA}

\author[0000-0002-0885-7215]{Mark~E.~Everett}
\affiliation{U.S. National Science Foundation National Optical-Infrared Astronomy Research Laboratory, 950 N.\ Cherry Ave., Tucson, AZ 85719, USA}

\author[0000-0001-6545-639X]{Eric B.\ Ford}
\affiliation{\PSUAA}
\affiliation{\PSUCEHW}
\affiliation{Institute for Computational and Data Sciences,  Penn State University, University Park, PA, 16802, USA}
\affiliation{Center for Astrostatistics,  525 Davey Laboratory, 251 Pollock Road, University Park, PA, 16802, USA}

\author[0000-0003-1312-9391]{Samuel Halverson}
\affiliation{Jet Propulsion Laboratory, California Institute of Technology, 4800 Oak Grove Drive, Pasadena, California 91109}

\author[0000-0002-3985-8528]{Jesus~Higuera}
\affiliation{U.S. National Science Foundation National Optical-Infrared Astronomy Research Laboratory, 950 N.\ Cherry Ave., Tucson, AZ 85719, USA}

\author[0000-0002-4475-4176]{Henry A. Kobulnicky}
\affiliation{Department of Physics \& Astronomy, University of Wyoming, Laramie, WY 82070, USA}

\author[0000-0001-9626-0613]{Daniel M. Krolikowski}
\affiliation{Steward Observatory, The University of Arizona, 933 N. Cherry Avenue, Tucson, AZ 85721, USA}

\author[0000-0002-2401-8411]{Alexander Larsen}
\affiliation{Department of Physics \& Astronomy, University of Wyoming, Laramie, WY 82070, USA}

\author[0000-0002-9082-6337]{Andrea S.J. Lin}
\affiliation{Department of Astronomy, California Institute of Technology, 1200 E California Blvd, Pasadena, CA 91125, USA}

\author[0000-0001-9596-7983]{Suvrath Mahadevan}
\affiliation{\PSUAA}
\affiliation{\PSUCEHW}

\author[0000-0003-0241-8956]{Michael W. McElwain}
\affiliation{NASA Goddard Space Flight Center, 8800 Greenbelt Road, Greenbelt, MD 20771, USA}

\author[0000-0002-0048-2586]{Andrew Monson}
\affiliation{Steward Observatory, University of Arizona, 933 N. Cherry Ave, Tucson, AZ 85721, USA}

\author[0000-0001-8720-5612]{Joe P. Ninan}
\affiliation{Department of Astronomy and Astrophysics, Tata Institute of Fundamental Research, Homi Bhabha Road, Colaba, Mumbai 400005, India}

\author[0000-0003-1324-0495]{Leonardo A. Paredes}
\affiliation{Steward Observatory, University of Arizona, 933 N. Cherry Ave, Tucson, AZ 85721, USA}

\author[0009-0004-7817-2547]{Yatrik~G.~Patel}
\affiliation{U.S. National Science Foundation National Optical-Infrared Astronomy Research Laboratory, 950 N.\ Cherry Ave., Tucson, AZ 85719, USA}

\author[0000-0003-0149-9678]{Paul Robertson}
\affiliation{Department of Physics \& Astronomy, The University of California, Irvine, Irvine, CA 92697, USA}

\author[0009-0006-7023-1199]{Gabrielle Ross}
\affiliation{Department of Astrophysical Sciences, Princeton University, 4 Ivy Lane, Princeton, NJ 08540, USA}

\author[0000-0001-8127-5775]{Arpita Roy}
\affiliation{Astrophysics \& Space Institute, Schmidt Sciences, New York, NY 10011, USA}

\author[0000-0002-4046-987X]{Christian Schwab}
\affiliation{School of Mathematical and Physical Sciences, Macquarie University, Balaclava Road, North Ryde, NSW 2109, Australia}

\author[0000-0001-7409-5688]{Gudmundur Stefansson}
\affiliation{Anton Pannekoek Institute for Astronomy, University of Amsterdam, Science Park 904, 1098 XH Amsterdam, The Netherlands}

\author[0000-0002-5951-8328]{Daniel J. Stevens}
\affiliation{Department of Physics \& Astronomy, University of Minnesota Duluth, Duluth, MN 55812, USA}

\author[0000-0001-7246-5438]{Andrew M. Vanderburg}
\altaffiliation{Sloan Fellow}
\affiliation{Department of Physics and Kavli Institute for Astrophysics and Space Research, Massachusetts Institute of Technology, Cambridge, MA 02139, USA}
 
\author[0000-0001-6160-5888]{Jason Wright}
\affiliation{\PSUAA}
\affiliation{\PSUCEHW}
\affiliation{\PSETI}

\begin{abstract}
We present the confirmation of TOI-5573\,b, a Saturn-sized exoplanet on an 8.79\,days orbit around an early M-dwarf (3790\,K, 0.59\,\Rsun, 0.61\,\Msun, 12.30\,Jmag). TOI-5573\,b has a mass of $112^{+18}_{-19}$\,\Mearth (0.35$\pm$0.06\,\Mjup) and a radius of 9.75$\pm$0.47\,\Rearth (0.87$\pm$0.04\,\Rjup), resulting in a density of $0.66^{+0.16}_{-0.13}$ g/cm³, akin to that of Saturn. The planet was initially discovered by TESS and confirmed using a combination of 11 transits from four TESS Sectors (20, 21, 47, and 74), ground-based photometry from the Red Buttes Observatory, and high-precision radial velocity data from the Habitable-zone Planet Finder (HPF) and NEID spectrographs, achieving a 5$\sigma$ precision on the planet’s mass. TOI-5573 b is one of the coolest Saturn-like exoplanets discovered around an M-dwarf, with an equilibrium temperature of only 528 $\pm$ 10\,K, making it a valuable target for atmospheric characterization. Saturn-like exoplanets around M-dwarfs likely form through core accretion, with increased disk opacity slowing gas accretion and limiting their mass. The host star’s super-solar metallicity supports core accretion, but uncertainties in M-dwarf metallicity estimates complicate definitive conclusions. Compared to other GEMS (Giant Exoplanets around M-dwarf Stars) orbiting metal-rich stars, TOI-5573\,b aligns with the observed pattern that giant planets preferentially form around M-dwarfs with super-solar metallicity. Further high-resolution spectroscopic observations are needed to explore the role of stellar metallicity in shaping the formation and properties of giant exoplanets like TOI-5573\,b.
\end{abstract}

\section{Introduction} \label{sec:intro}
Giant Exoplanets around M-dwarf Stars, hereafter GEMS, are a rare planet population that are characterized by their large radii, typically between 8-15\,\Rearth, and M$_p$ sin(i) $>$ 80\,\Mearth. The rarity of GEMS make them stand out from the more common, smaller planets typically found around M-dwarfs. The formation mechanisms behind these rare giants remain an active area of research, with current planet formation models struggling to explain their presence around low-mass M-dwarf stars \citep{Burn2021}. Their discovery challenges existing theories and invites further exploration into the physical processes that could enable such large planets to form in these low-mass environments.

Core accretion theory \citep{laughlin2004,idalin2005} predicts a low occurrence rate for GEMS. This is because the formation of giant planets is challenging around M-dwarfs due to the smaller median masses of their protoplanetary disks, which are typically correlated with the low mass of their host stars. These disks may not contain enough material to accrete the large cores necessary for giant planet formation, resulting in the scarcity of GEMS. In particular, the time required for the core to accumulate enough mass in order to undergo runaway gas envelope accretion in these disks might exceed the typical lifetime of the disk, making the formation of such large planets even more improbable \citep{Burn2021}. Thus, core accretion models suggest that any GEMS that do exist are the result of either unusually massive disks or particularly favorable disk environments \citep{Pan2024}. We note that there is no fundamental problem forming giants around early M dwarfs in general; the main challenges arise for stars with M$* \lesssim 0.5$M$\odot$.

Another proposed formation mechanism is gravitational instability \citep{Boss2006}, which suggests that giant planets could form through the rapid collapse of material in a protoplanetary disk at large distances from the host star. However, this scenario requires a relatively high disk-to-star mass ratio of about 10\% \citep{BossKanodia2023}, further contributing to the predicted low occurrence rate of GEMS. For this mechanism to work, the outer regions of the disk must cool rapidly enough to allow the material to collapse under its own gravity, a condition that is rarely met, especially around M-dwarfs with their comparatively low-mass disks. Indeed, radial velocity (RV) studies conducted by \cite{endl2006, kovacs2013, sabotta2021, pinamonti2022} have been able to set only an upper limit on the occurrence rate of these planets, ranging from 1\% to 3\%. These low occurrence rates indicate that while gravitational instability might contribute to GEMS formation in some cases, it is likely not prolific at doing so.

Despite these challenges, the nearly all-sky survey conducted by the Transiting Exoplanet Survey Satellite (TESS; \citealt{ricker2015_tess}) has made significant strides in the discovery of transiting GEMS. This mission has uncovered dozens of new GEMS, significantly expanding the known sample. Notable recent discoveries include, but are not limited to, planets such as TOI-5344,b \citep{Han2024}, TOI-5688,A,b \citep{Reji2024}, Kepler-45,b \citep{Johnson2012}, TOI-762,A,b \citep{Hartmann2024}, TOI-3984,A,b \citep{Canas2023}, HATS-71,b \citep{bakos2020}, and TOI-3235,b \citep{Hobson2023}. TESS’s ability to observe large portions of the sky for extended periods allows it to detect transits of planets that might otherwise go unnoticed. For instance, \cite{gan2023} reported an occurrence rate of 0.27$\pm$0.09\% for hot Jupiters with radii between 7 and 22\,\Rearth (or 0.06 and 2\,\Rjup) and orbital periods between 0.8 and 10\,days around early-type M-dwarfs (0.45\,\Msun$\leq$M$\leq$0.65\,\Msun). Meanwhile, \cite{bryant2023} found an even lower occurrence rate of 0.194$\pm$0.072\% across the entire range of M-dwarf stars ($<$0.71\,\Msun). These studies highlight the power of TESS to uncover new GEMS and provide valuable statistical insights into their occurrence rates, even though such planets remain rare.

Accurate mass measurements of transiting GEMS are essential for (i) confirming their planetary nature, (ii) investigating the relationships between planetary mass and other stellar or planetary properties, and (iii) determining their atmospheric compositions, which provide crucial clues about their formation history. These measurements, typically obtained through RV observations, offer key insights into the formation and evolution of these rare giants around M-dwarf stars \citep{MullerHelled2024, Kanodia2024}. In particular, knowing the planet’s mass enables calculation of the atmospheric scale height, which in turn allows constraints on the atmospheric metallicity (e.g., \citealt{Canas2025}). This can be compared to the planet’s bulk metallicity to infer the relative contributions of gas and solids during formation, providing insight into the protoplanetary disk’s mass budget and accretion processes \citep{Kanodia2024}. Additionally, they help distinguish GEMS from brown dwarfs and low-mass stars, which can produce similar transit signals and lead to astrophysical false positives, such as eclipsing binaries.

Despite their importance, the current sample of transiting GEMS remains small ($\sim$30), limiting our ability to fully characterize their properties. To address this, we have launched the volume-limited ($<$200 pc) \textit{Searching for GEMS} survey \citep{Kanodia2024_motivation}, aiming to expand the sample to $\sim$40 transiting GEMS. =This increase will enable robust statistical comparisons with their FGK-star counterparts \citep{Kanodia2024_motivation}, offering new insights into their formation mechanisms and the unique conditions of planet formation around low-mass stars.

Here, as part of the \textit{Searching for GEMS} survey, we present the discovery of the transiting Saturn-like exoplanet TOI-5573\,b around an M-dwarf, utilizing photometry from four TESS sectors, ground-based observations from the 0.6\,m telescope at Red Buttes Observatory (RBO), as well as high-resolution spectra and precise radial velocity (RV) data from the Habitable-zone Planet Finder (HPF) and NEID spectrographs. In Section \ref{sec:observations} we detail the observational data used in this analysis and, in Section \ref{sec:stellar} we focus on the determination of stellar properties of the host star. In Section \ref{sec:joint}, we outline the joint analysis of transit and RV data through Bayesian methods. Finally, in Section \ref{sec:discussion}, we place TOI-5573\,b in context with other GEMS systems, and provide a qualitative discussion of the planet's formation mechanism. 

\section{Observations}\label{sec:observations}
\begin{figure*}[!htbp]
    \centering
    \includegraphics[width=\linewidth]{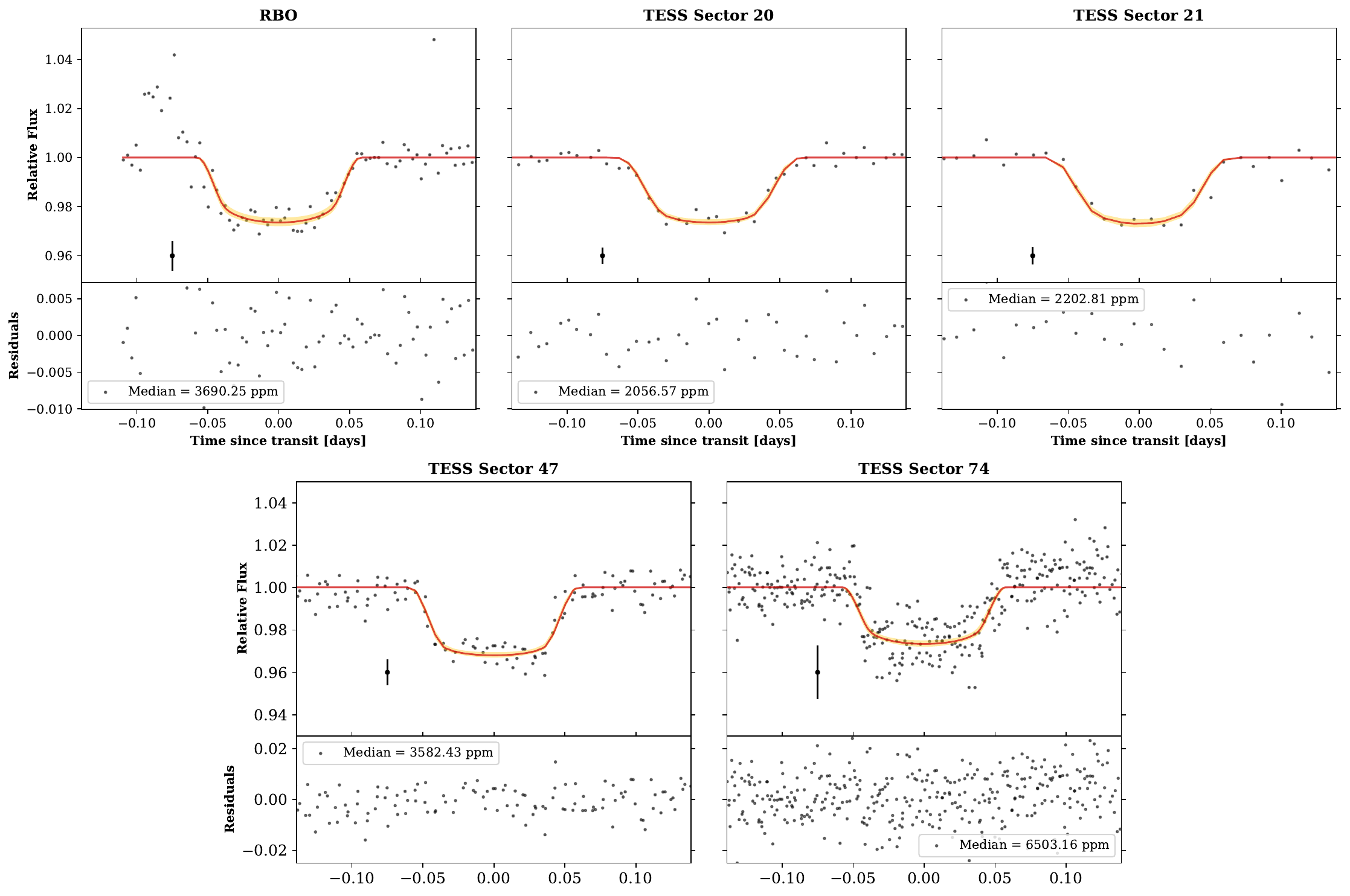}
    \caption{The raw (gray) phase-folded light curves of RBO and \texttt{TGLC} TESS sectors 20, 21, 47, and 74 photometry for TOI-5573. The best joint fit model is shown in red, and the 1$\sigma$ confidence intervals are shown in yellow.}
    \label{fig:transits}
\end{figure*}

\begin{figure*}[!htbp]
    \centering
    \includegraphics[width=\linewidth]{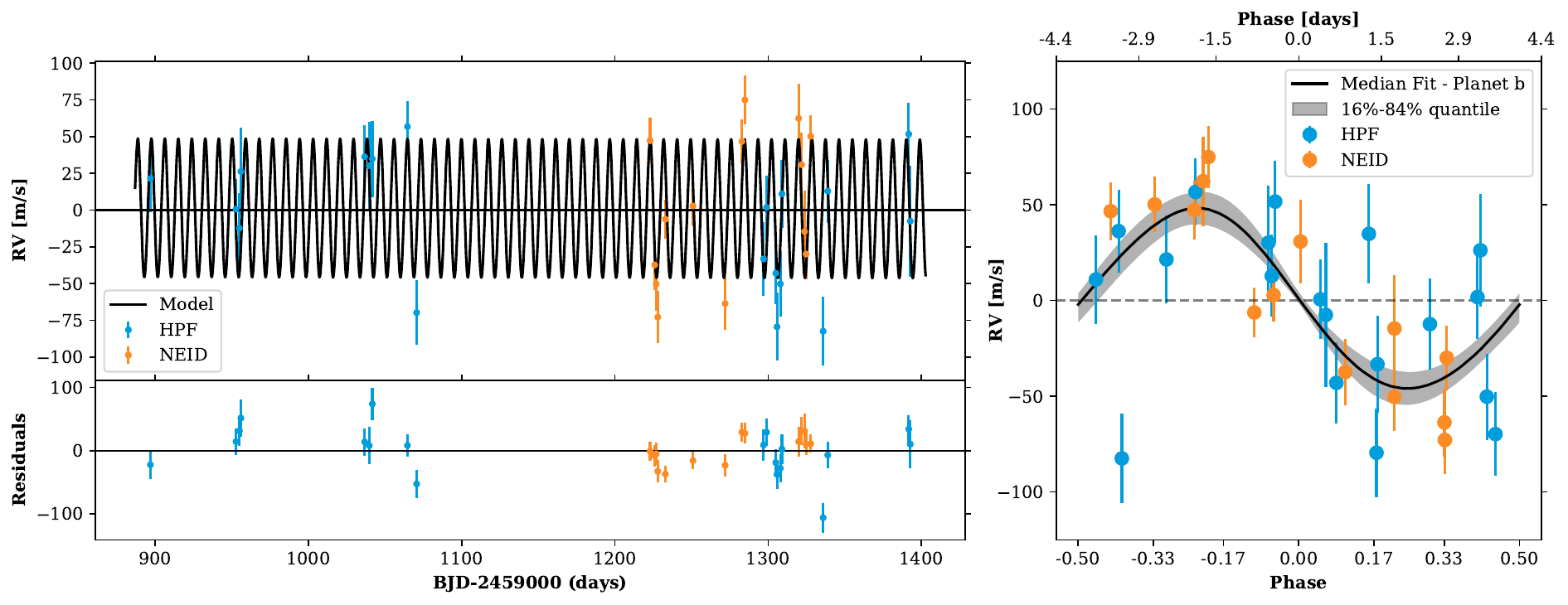}
    \caption{The binned HPF (blue) and NEID (orange) RV measurements of TOI-5573. Left: time series binned by each night. Right: phase-folded RV (black). The best fit model is shown in black, and the 1$\sigma$ confidence intervals are shown in gray.}
    \label{fig:rvs}
\end{figure*}

The best-fit transit model to the data is shown in a phase-folded plot in Figure~\ref{fig:transits}, while the best-fit RV model and data are plotted in Figure~\ref{fig:rvs} both as a time series and phase-folded. 

\subsection{TESS}\label{sec:TESS}
TOI-5573 (TIC 459762279; Gaia DR3 1023967108706495360) was observed with TESS in Sectors 20 and 21 from 2019 December 24 to 2020 February 18 at 1800s cadence, in Sector 47 from 2021 December 31 to 2022 January 27 at 600s cadence, and in Sector 74 from 2024 January 3 to 2024 January 30 at 200s cadence, all within the full-frame images (FFIs). During these observations, TESS detected a total of 11 transits. The planet candidate TOI-5573\,b was identified by the TESS Faint Star Search \citep{kunimoto2022_faint} using data from the Quick Look Pipeline (QLP; \citealt{huang2020_qlp, kunimoto2022_yield}). 

For our analysis, we use the TESS-Gaia Light Curve (\texttt{TGLC}; \citealt{han2023_tglc}) for all four sectors. \texttt{TGLC} employs locally-modeled point spread functions to simulate and remove contamination of nearby stars, correcting light curve dilutions and improving photometric precision. We run the Python package \texttt{TGLC} on FFI cuts of $90 \times 90$ pixels and extract the calibrated aperture light curves (\texttt{cal\_aper\_flux}). The calibrated aperture light curves are derived from the contamination-removed FFI with a $3 \times 3$ aperture and detrended with the `biweight' method from the Python package \texttt{wotan} \citep{Hippke2019}.

\subsection{Ground based photometric follow up}\label{sec:photometry}
\subsubsection{RBO 0.6m Telescope}
We observed a transit of TOI-5573\,b on the night of 2023 May 1, using the Red Buttes Observatory 0.6\,m telescope (RBO; \citealt{kasper_remote_2016}). The observation was conducted with a slight defocus to achieve a PSF FWHM of 2.0\arcsecond, with exposure times of 240\,s. The transit was observed in the Bessell I filter with the air mass ranging from 1.07 to 2.7 throughout the observation. Observing notes indicate that clouds were present at the beginning of the transit, which likely explains the poor data quality in that region. The raw data were reduced using a custom \texttt{python} differential aperture photometry pipeline, based on the methodology described in \citet{monson2017}, as illustrated in Figure~\ref{fig:transits}.

\subsubsection{Zwicky Transient Facility (ZTF)}
ZTF operates at the 1.2\,m telescope at Palomar Observatory, California, and observes the entire Northern sky roughly every 2\,days to search for transient events \citep{Masci2019}. We acquired ZTF data from their publicly accessible API through the \texttt{ztfquery} module \citep{Rigault2018}. ZTF uses all three of its filters (green, red, and infrared) with the same 30-second exposure time to observe all of its targets across the Northern sky. We downloaded and analyzed the ZTF data with the DEATHSTAR pipeline \citep{Ross2024}. DEATHSTAR takes in the TIC ID of the star, downloads the ZTF data, extracts the light curves of each star in the field from the ZTF images through a custom 2-D gaussian fit, and plots the light curves for quick manual verification. Using DEATHSTAR, we confirmed this was an on-target detection. We estimate the median PSF FWHM to be 0.302\arcsecond. This number is the Gaussian standard deviation of the PSF width, which can be converted to FWHM (as it is more often used to characterize the seeing disk) by multiplying by a factor of 2.3. Given the cadence, the data is not utilized during the joint fitting (Section~\ref{sec:joint}).


\subsection{High Contrast Imaging}
\begin{figure}[!htbp]
    \centering
    \includegraphics[width=\linewidth]{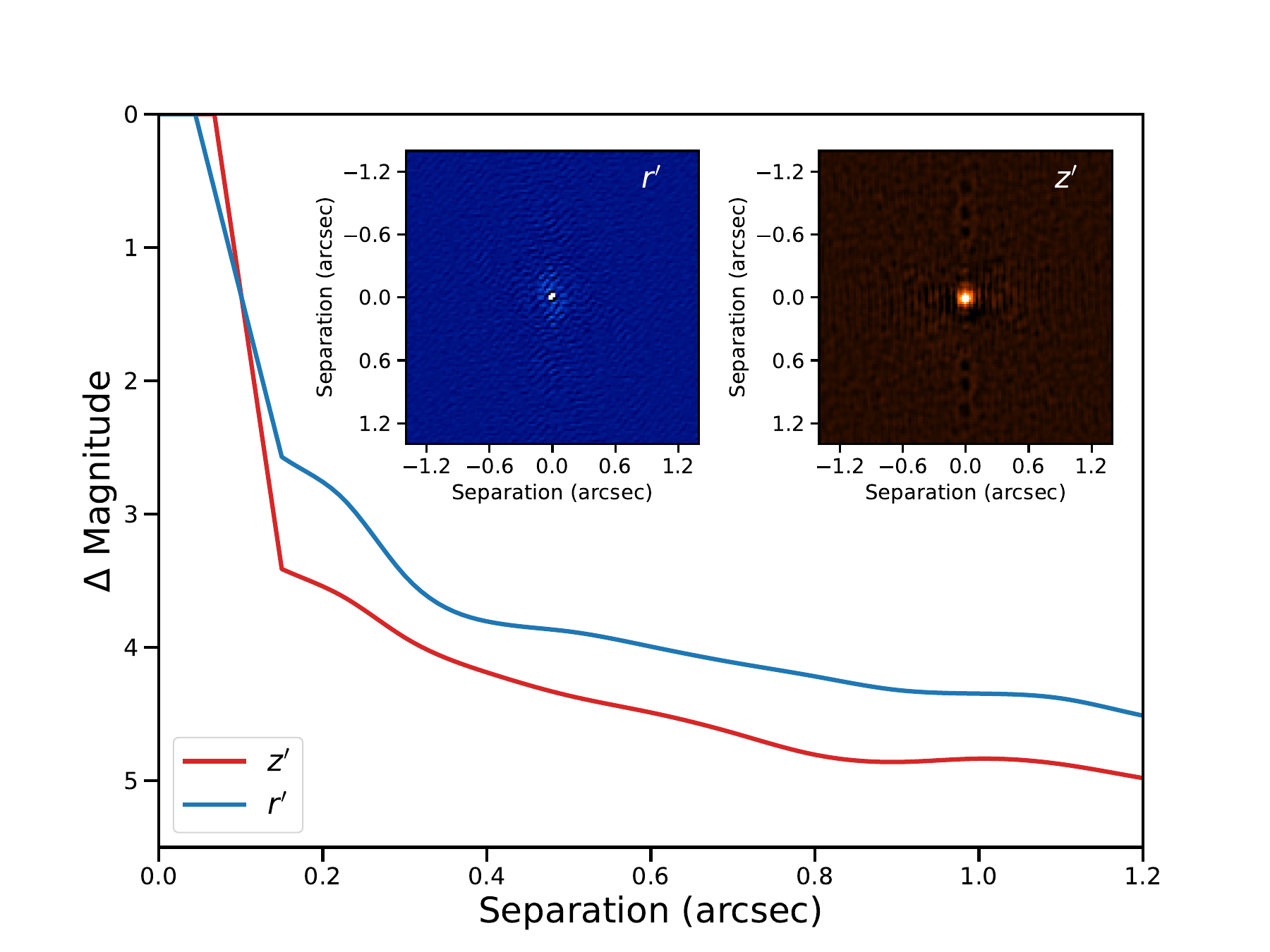}
    \caption{NESSI Speckle Imaging in r' and z' bands in the inset 2.4\arcsecond across. The curve shows the 5-$\sigma$ contrast curve for TOI-5573 in both $z$ and $r$ bands. The contrast curves indicate that there are no bright companions within 1.2\arcsecond from the host star.}
    \label{fig:speckle}
\end{figure}

We observed TOI-5573 on the night of 2023 January 27 using the NN-Explore Exoplanet Stellar Speckle Imager \citep[NESSI;][]{Scott2018_speckimg} attached to the WIYN 3.5\,m telescope at Kitt Peak National Observatory. Our goal was to detect faint background stars and nearby stellar companions. We captured a series of 9000 40\,ms diffraction-limited exposures spanning 6 minutes, utilizing the SDSS $r^\prime$ and SDSS $z^\prime$ filters on NESSI. The speckle images were processed according to the methods described in \cite{Howell2011}. No stellar sources were detected below a magnitude limit of $\Delta r'$ = 3.88\,mag at 0.5\arcsecond and 4.51\,mag at 1.2\arcsecond, or $\Delta z'$ =  4.36\,mag at 0.5\arcsecond and 4.98\,mag at 1.2\arcsecond, as shown in Figure~\ref{fig:speckle}.

\subsection{Radial velocity follow-up with HPF}\label{sec:hpfrvs}

TOI-5573 was observed with the near-infrared Habitable-zone Planet Finder (HPF; \citealt{mahadevan_habitable-zone_2012,mahadevan_habitable-zone_2014,kanodia_overview_2018}) from 2022 November 13 to 2024 March 23. HPF is situated at the 10\,m Hobby–Eberly Telescope (HET; \citealt{ramsey_early_1998, Hill2021}). We processed the raw HPF data using \texttt{HxRGproc} algorithms \citep{ninan_habitable-zone_2018} and follow the method described in \citet{stefansson_sub-neptune-sized_2020} for deriving an interpolated wavelength solution to avoid cross-contamination from simultaneous calibration light. RVs were then derived from the spectra using a modified version of the \texttt{SpEctrum Radial Velocity AnaLyser} pipeline (SERVAL; \citealt{zechmeister_spectrum_2018}), which is described further in \cite{Stefansson2023}. Finally, we applied barycentric correction to the individual spectra using \texttt{barycorrpy}, a Python implementation of the algorithms from \citet{wright_eastman_2014}, developed by \citet{kanodia_wright_2018}.

We obtained a total of 38 exposures of TOI-5573. The observations spanned 19 nights, with two 15 min exposures binned together per visit taken on each night. The median signal-to-noise ratio (S/N) was 40 at $\sim$1070\,nm, and the median uncertainty ($\sigma$) was 23\,m/s. The final HPF RVs are detailed in Table~\ref{tab:rvs} and illustrated in Figure~\ref{fig:rvs}.

\subsection{Radial velocity follow-up with NEID}\label{sec:neidrvs}
NEID \citep{Schwab2016_NEID, Halverson2016_NEID_errorbudget} is an ultra-stabilized, high-resolution (R$\sim$110,000), red-optical (380--930\,nm) extreme-precision RV spectrograph on the WIYN 3.5\,m Telescope at Kitt Peak National Observatory\footnote{The WIYN Observatory is a joint facility of the NSF's National Optical-Infrared Astronomy Research Laboratory, Indiana University, the University of Wisconsin-Madison, Pennsylvania State University, Purdue University and Princeton University.}. We obtained 14 visits of TOI-5573 with NEID, consisting of 14 individual exposures. As with HPF, we forego simultaneous calibration and instead rely on the interpolated wavelength solution method.

NEID data are automatically processed by the NEID Data Reduction Pipeline (DRP)\footnote{\url{https://neid.ipac.caltech.edu/docs/NEID-DRP/}}. Instead of using the DRP-calculated cross-correlation RVs, we retrieve the calibrated 1D spectra from the NEID Data Archive \texttt{v1.3}\footnote{\url{https://neid.ipac.caltech.edu/search.php}} and use a NEID-adapted version of \texttt{SERVAL} to derive RVs via template-matching \citep{Stefansson2022_GJ3470b}. These RVs are calculated from the central 7000 pixels of 44 orders centered between 5070--8900\,$\text{\AA}$ (order indices 52--104). The median S/N was 5 at 850\,nm, and the median uncertainty ($\sigma$) was 23\,m/s. The final NEID RVs are detailed in Table~\ref{tab:rvs} and illustrated in Figure~\ref{fig:rvs}.

\begin{table}[htbp]
\centering
\begin{tabular}{cccc}
\hline
\hline
BJD\textsubscript{TDB}(days) & RV(m s\textsuperscript{-1}) & $\sigma$ (m s\textsuperscript{-1}) & Instrument\\[0pt]
\hline
2459896.93154 & 35.73 & 22.87 & HPF \\[0pt]
2459952.78553 & 14.89 & 20.74 & HPF \\[0pt]
2459954.96852 & 1.97 & 23.99 & HPF \\[0pt]
2459955.97476 & 40.51 & 29.59 & HPF \\[0pt]
2460036.73971 & 50.53 & 21.69 & HPF \\[0pt]
2460039.72592 & 44.50 & 29.56 & HPF \\[0pt]
2460041.72772 & 49.09 & 25.84 & HPF \\[0pt]
2460064.66553 & 71.01 & 17.55 & HPF \\[0pt]
2460070.64370 & -55.49 & 21.83 & HPF \\[0pt]
2460297.02819 & -18.97 & 25.21 & HPF \\[0pt]
2460299.02343 & 16.08 & 21.67 & HPF \\[0pt]
2460305.00762 & -28.72 & 21.19 & HPF \\[0pt]
2460305.80711 & -65.23 & 23.13 & HPF \\[0pt]
2460308.00592 & -35.95 & 22.52 & HPF \\[0pt]
2460309.01238 & 25.29 & 23.08 & HPF \\[0pt]
2460335.92177 & -68.20 & 23.49 & HPF \\[0pt]
2460338.91078 & 27.14 & 21.46 & HPF \\[0pt]
2460391.76402 & 65.95 & 21.44 & HPF \\[0pt]
2460392.77328 & 6.77 & 37.74 & HPF \\[0pt]
\hline
2460222.99794 & 74.50 & 15.44 & NEID\\[0pt]
2460226.00662 & -10.07 & 17.26 & NEID\\[0pt]
2460226.98482 & -23.03 & 18.01 & NEID\\[0pt]
2460227.99125 & -45.55 & 17.68 & NEID\\[0pt]
2460232.99228 & 21.02 & 13.02 & NEID\\[0pt]
2460250.97054 & 30.14 & 13.68 & NEID\\[0pt]
2460271.97140 & -36.37 & 17.78 & NEID\\[0pt]
2460282.91351 & 73.95 & 15.24 & NEID\\[0pt]
2460284.85987 & 102.24 & 16.42 & NEID\\[0pt]
2460319.94666 & 89.64 & 23.38 & NEID\\[0pt]
2460321.88899 & 58.10 & 21.89 & NEID\\[0pt]
2460323.76032 & 12.67 & 27.67 & NEID\\[0pt]
2460324.80297 & -2.64 & 16.97 & NEID\\[0pt]
2460327.77877 & 77.50 & 14.45 & NEID\\[0pt]
\hline
\end{tabular}
\caption{\label{tab:rvs}RV estimates of TOI-5573, taken with HPF and NEID. The RV values from HPF are binned down to one day.}
\end{table}

\section{Stellar Parameters}\label{sec:stellar}
\subsection{HPF-SpecMatch}
\begin{figure*}[!htbp]
    \centering
    \includegraphics[width=\linewidth]{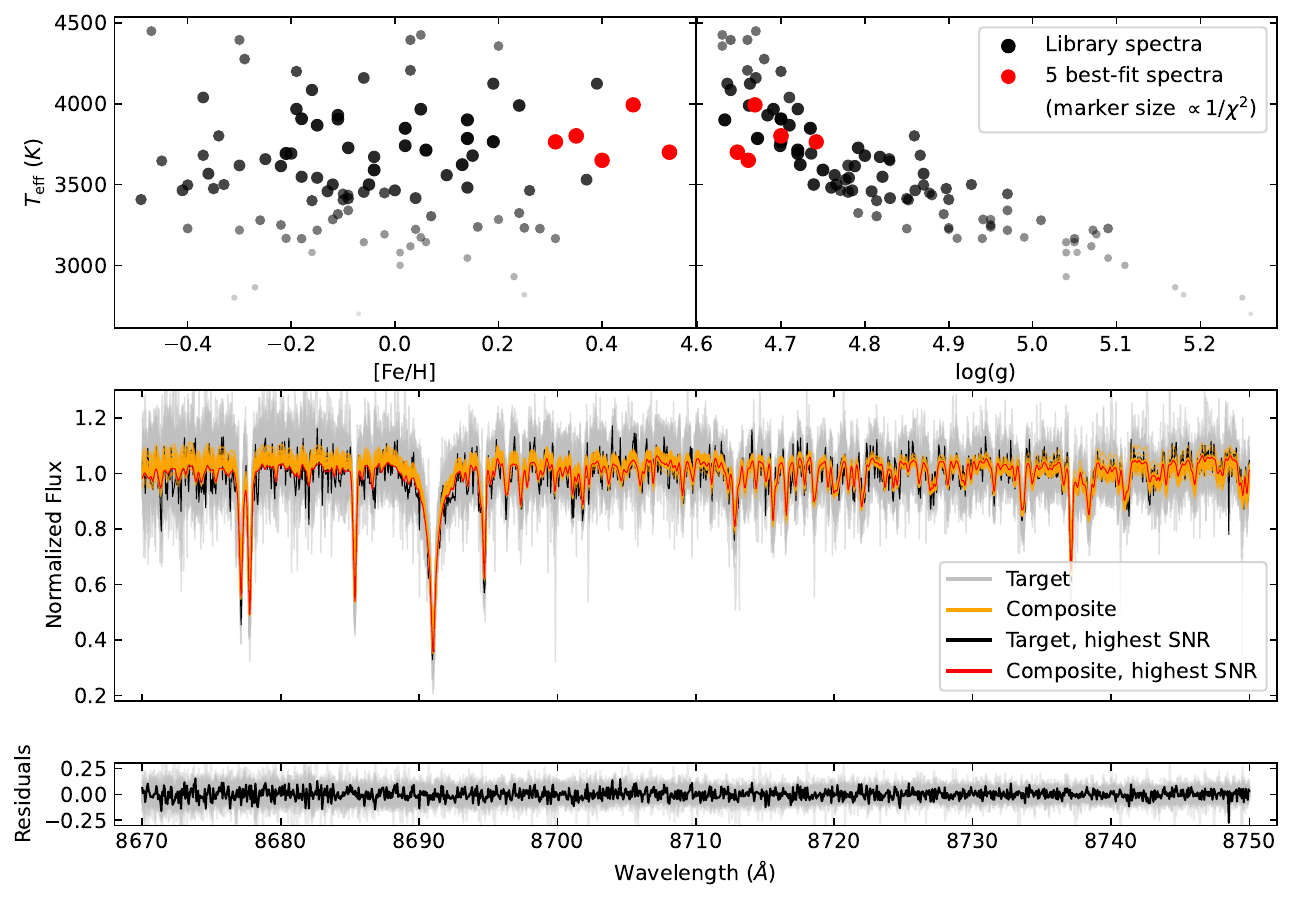}
    \caption{The \texttt{HPF-SpecMatch} spectra fit results for order index 5. Top: two plots showing the five best-fit stars (red) selected to compose the spectra. The size and transparency of these five stars and the other library stars (black) are inversely proportional to the calculated $\chi^2$ initial value where we compare TOI-5573 spectra to the library star spectra. Larger and darker points signify a lower $\chi^2$ initial value, implying a better fit to the target spectrum. Middle: all HPF order index 5 spectra of TOI-5573 (gray) and their best-fit composite (yellow). The highest S/N spectrum is shown in black and its best-fit composite is in red. Bottom: the residuals of the best-fit composites for all spectra (gray) and for the highest S/N spectrum (black). The featureless residuals suggest a high-quality fit.}
    \label{fig:hpf-specmatch}
\end{figure*}

We utilized the \texttt{HPF-SpecMatch} package \citep{stefansson_sub-neptune-sized_2020} to estimate stellar parameters from HPF spectra using a two-step $\chi^2$-based algorithm (see Figure~\ref{fig:hpf-specmatch}). This method identifies the best-matching library stars for the target spectrum, using a library of 100 stars that span $2700\mathrm{K} \le T_{e} \le 4500~\mathrm{K}$, $4.6<\log g_\star < 5.3$, and $-0.5 < \mathrm{[Fe/H]} < 0.5$. The target spectrum is compared against all library spectra, and the top five best-fit stars are then used to create a composite spectrum that closely aligns with the target (see Figure~\ref{fig:hpf-specmatch}).

For TOI-5573, spectral matching was performed using HPF order index 5 (8656-8767\,\AA) due to its minimal telluric contamination. No significant rotational broadening was detected, resulting in an upper limit of \vsini $<$2\,km/s for TOI-5573, which is consistent with TESS photometry that shows no detectable rotational modulation. The derived effective temperature, metallicity, and surface gravity for TOI-5573 are listed in Table~\ref{tab:stellarparam}. Notably, the metallicity value of [Fe/H] = 0.42$\pm$0.16 is near the upper limit of the \texttt{HPF-SpecMatch} library, raising the question of whether TOI-5573 might have a metallicity higher than 0.5. However, even though the top five matching stars (GJ 205, GJ 134, GJ 96, BD+29 2279, and GJ 895) are among the highest metallicity in the library, we have adopted the metallicity provided by \texttt{HPF-SpecMatch}.  Additionally, we estimated the metallicity using a photometric calibration with \texttt{METaMorPHosis} \citep{Duque-Arribas2022}, yielding [Fe/H] = 0.37$\pm$0.21. Despite the significant uncertainties, both methods consistently indicate that TOI-5573 has an [Fe/H] considerably higher than Solar. However, we advise caution in interpreting this beyond being a super-solar metallicity M-dwarf.

\subsection{Model-dependent Stellar Parameters}
\begin{figure}[!htbp]
    \centering
    \includegraphics[width=\linewidth]{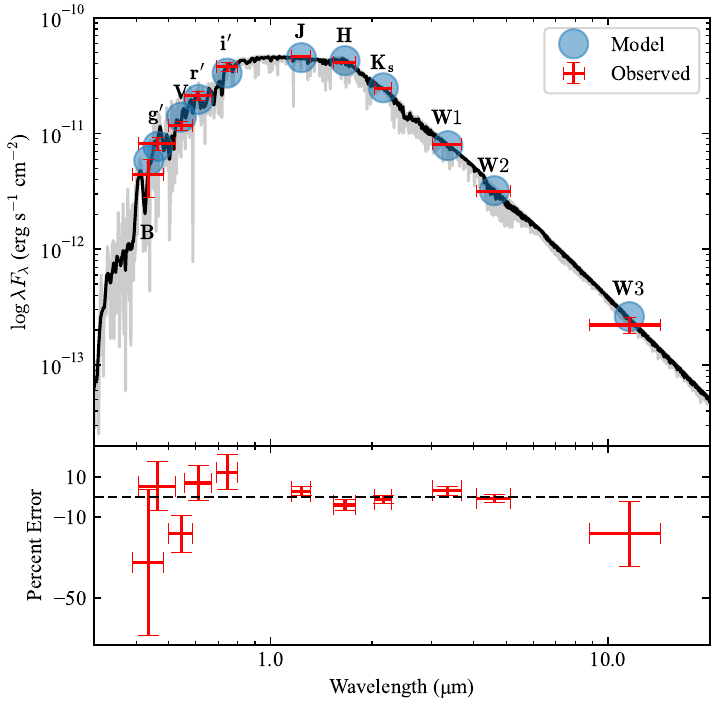}
    \caption{SED of TOI-5573 showing the broadband photometric measurements from Table~\ref{tab:stellarparam} in red (the x-error bar is the bandpass width) and the derived MIST model fluxes (as blue circles). A NextGen BT-SETTL spectrum \citep{allard2012} is overlaid for reference in gray (a smoothed version in black); the model spectrum is not used when fitting the SED.}
    \label{fig:sed}
\end{figure}

We used the \texttt{EXOFASTv2} package \citep{eastman2019} to model the spectral energy distribution (SED) of TOI-5573, which provided us with model-dependent stellar parameters such as stellar mass, radius, luminosity, and age. We utilized the default Modules for Experiments in Stellar Astrophysics Isochrones and Stellar Tracks (MIST) model grids \citep{choi2016,dotter2016} and applied Gaussian priors on reliable broadband photometry from APASS, 2MASS, and WISE, as well as on the spectroscopic stellar parameters from \texttt{HPF-SpecMatch} and the extinction-corrected geometric distance from \citep{bailerjones2021_gdr3}. The SED fit is shown in Figure~\ref{fig:sed}, and the derived model-dependent stellar parameters are detailed in Table~\ref{tab:stellarparam}. 

\begin{deluxetable*}{lccc}
\tablecaption{Summary of stellar parameters for TOI-5573. \label{tab:stellarparam}}
\tablehead{\colhead{~~~Parameter}&  \colhead{Description}&
\colhead{Value}&
\colhead{Reference}}
\startdata
\multicolumn{4}{l}{\hspace{-0.2cm} Main identifiers:}  \\
~~~TOI & \tess Object of Interest & 5573 & \tess mission \\
~~~TIC & \tess Input Catalog  & 459762279 & Stassun \\
~~~2MASS & $\cdots$ & J09054223+5518267 & 2MASS  \\
~~~Gaia DR3 & $\cdots$ & 1023967108706495360  & Gaia DR3\\
\multicolumn{4}{l}{\hspace{-0.2cm} Equatorial Coordinates, Proper Motion and Spectral Type:} \\
~~~$\alpha_{\mathrm{ICRS, 2015.5}}$ &  Right Ascension (RA) & 09:05:42.24 & Gaia DR3\\
~~~$\delta_{\mathrm{ICRS, 2015.5}}$ &  Declination (Dec) & +55:18:25.67 & Gaia DR3\\
~~~$\mu_{\alpha}$ &  Proper motion (RA, \unit{mas/yr}) &  -3.63332 & Gaia DR3\\
~~~$\mu_{\delta}$ &  Proper motion (Dec, \unit{mas/yr}) & -56.9802 & Gaia DR3 \\
~~~$d$ &  Distance in pc  & $185.8^{+1.1}_{-1.3}$ & Bailer-Jones \\
\multicolumn{4}{l}{\hspace{-0.2cm} Optical and near-infrared magnitudes:}  \\
~~~$B$ & Johnson B mag & $16.991\pm0.393$ & APASS\\
~~~$V$ & Johnson V mag & $15.574\pm0.1$ & APASS\\
~~~$g^{\prime}$ &  Sloan $g^{\prime}$ mag  & $16.144\pm0.132$ & APASS\\
~~~$r^{\prime}$ &  Sloan $r^{\prime}$ mag  & $14.81\pm0.092$ & APASS \\
~~~$i^{\prime}$ &  Sloan $i^{\prime}$ mag  & $13.969\pm0.094$ & APASS \\
~~~$J$ & $J$ mag & $12.305\pm0.024$ & 2MASS\\
~~~$H$ & $H$ mag & $11.646\pm0.028$ & 2MASS\\
~~~$K_s$ & $K_s$ mag & $11.448\pm0.022$ & 2MASS\\
~~~$W1$ &  WISE1 mag & $11.343\pm0.023$ & WISE\\
~~~$W2$ &  WISE2 mag & $11.374\pm0.021$ & WISE\\
~~~$W3$ &  WISE3 mag & $11.434\pm0.174$ & WISE\\
\multicolumn{4}{l}{\hspace{-0.2cm} Spectroscopic Parameters$^a$:}\\
~~~$T_{\mathrm{eff}}$ &  Effective temperature in \unit{K} & $3790\pm59$ & This work\\
~~~$\mathrm{[Fe/H]}$ &  Metallicity in dex & $0.42\pm0.16$ & This work\\
~~~$\log(g)$ & Surface gravity in cgs units & $4.68\pm0.04$ & This work\\
\multicolumn{4}{l}{\hspace{-0.2cm} Model-Dependent Stellar SED and Isochrone fit Parameters$^b$:}\\
~~~$M_*$ &  Mass in $M_{\odot}$ & $0.619\pm0.023$ & This work\\
~~~$R_*$ &  Radius in $R_{\odot}$ & $0.594\pm0.011$ & This work\\
~~~$L_*$ &  Luminosity in $L_{\odot}$ & $0.0693^{+0.0025}_{-0.0023}$ & This work\\
~~~$\rho_*$ &  Density in $\unit{g/cm^{3}}$ & $4.08^{+0.26}_{-0.24}$ & This work\\
~~~Age & Age in Gyrs & $9.7^{+3.9}_{-2.7}$ & This work\\
~~~$A_v$ & Visual extinction in mag & $0.075^{+0.074}_{-0.053}$ & This work\\
\multicolumn{4}{l}{\hspace{-0.2cm} Other Stellar Parameters:}           \\
~~~\vsini &  Rotational velocity in \unit{km/s}  & $<$2 & This work\\
~~~$\Delta$RV &  ``Absolute'' radial velocity in \unit{km/s} & -17.581953  & Gaia DR3\\
\enddata
\tablenotetext{}{References are: Stassun \citep{stassun2018_tic}, 2MASS \citep{cutri2003_2mass}, Gaia DR2 \citep{gaiadr2_2018}, Bailer-Jones \citep{bailerjones2021_gdr3}, APASS \citep{henden2018_apass}, WISE \citep{wright2010_wise}}
\tablenotetext{a}{Derived using the HPF-SpecMatch algorithm from \cite{stefansson_sub-neptune-sized_2020}}
\tablenotetext{b}{{\tt EXOFASTv2} derived values using MIST isochrones with the \gaia parallax and spectroscopic parameters in $a$) as priors.}
\end{deluxetable*}

\section{Joint Fitting of Photometry and RVs}\label{sec:joint}
We performed a joint fit of photometry, and the NEID and HPF RVs (binned by night) using the \texttt{exoplanet} package \citep{exoplanet_dfm}, which is based on \texttt{PyMC3}, a Hamiltonian Monte Carlo package \citep{salvatier2016probabilistic}. The \texttt{exoplanet} package uses \texttt{starry} \citep{luger2019,agol2020} to simulate planetary transits, employing the analytical transit models by \citet{mandel_agol2002} along with a quadratic limb-darkening law. Limb-darkening priors in \texttt{exoplanet} are based on the reparameterization by \citet{kipping2023a} for uninformative sampling. Each transit (Figure~\ref{fig:transits}) is fitted independently with specific limb-darkening coefficients. We model the RV with the standard Keplerian model with free \textit{e} and $\omega$ while including offsets and jitter terms for HPF and NEID. Since the TESS light curves exhibit no rotational variability, we chose a model that does not incorporate the GP. 

We include the stellar mass, radius, temperature, transit mid-point, orbital period, and transit depth as priors in our posterior fits. We run four chains with 1500 burn-in steps followed by an additional 4500 steps to ensure robust planetary parameter estimation. To verify that our posteriors are well-mixed and independent, we assess convergence using the Gelman-Rubin statistic, where $\hat{R} \sim 1$ indicates strong convergence \citep{Ford2006}. We find that TOI-5573\,b is a Saturn-like planet with a radius of 9.75$\pm$0.47\,\Rearth and a mass of $112^{+18}_{-19}$\,\Mearth. and a density of $0.66^{+0.16}_{-0.13}$ g/$cm^3$. A summary of the inferred system parameters along with their corresponding confidence intervals is presented in Table~\ref{tab:orbitalparam}. 

\begin{deluxetable*}{llc}  
\tablecaption{Summary of orbital and physical parameters for TOI-5573\,b. \label{tab:orbitalparam}}
\tablehead{\colhead{~~~Parameter}&  \colhead{Units}  & \colhead{Value\textsuperscript{a}}}
\startdata
\sidehead{Orbital Parameters:}
Orbital Period & P (days) & $8.79758977^{+0.00000986}_{-0.00001001}$\\
Eccentricity & e & $0.072^{+0.067}_{-0.048}$\\
Argument of Periastron & $\omega$ (radians) & $0.5725^{+1.343}_{-1.753}$\\
Semi-amplitude Velocity & K (m/s) & $47.9^{+7.8}_{-8.1}$\\
Systemic Velocity\textsuperscript{b} & $\gamma_{\text{HPF}}, ~\gamma_{\text{NEID}}$ (m/s) & $14.2^{+9.5}_{-9.5},~ 27.2^{+7.4}_{-7.3}$\\
RV trend & $\frac{dv}{dt}$ (m/s) & $-0.51^{+4.72}_{-4.81}$\\
RV jitter & $\sigma_{\text{HPF}}, ~\sigma_{\text{NEID}}$ (m/s) & $34.32^{+10.45}_{-8.84},~19.46^{+9.35}_{-7.63}$\\
\hline
\sidehead{Transit Parameters:}
Transit Midpoint & T\textsubscript{0} (BJD\textsubscript{TDB}) & $2459600.45078^{+0.00061}_{-0.00063}$\\
Impact parameter & b & $0.43^{+0.10}_{-0.19}$ \\
Scaled Radius & $R_p/R_*$ & $0.150^{+0.005}_{-0.006}$\\
Scaled Semimajor Axis & $a/R_*$ & $25.74^{+0.72}_{-0.68}$\\
Orbital Inclination & i (degrees) & $89.01^{+0.39}_{-0.20}$\\
Transit Duration & T\textsubscript{14} (days) & $0.115^{+0.006}_{-0.004}$\\
Photometric Jitter\textsuperscript{c} & $\sigma_{TESS,S20}$ (ppm) & $17.97^{+115.30}_{-15.44}$\\
 & $\sigma_{TESS,S21}$ (ppm) & $18.25^{+116.35}_{-15.69}$\\
 & $\sigma_{TESS,S47}$ (ppm) & $15.03^{+87.15}_{-12.61}$\\
 & $\sigma_{TESS,S74}$ (ppm) & $15.96^{+92.47}_{-13.55}$\\
 & $\sigma_{RBO20221009}$ (ppm) & $33.32^{+323.13}_{-30.27}$\\
Dilution\textsuperscript{d} & $D_{TESS,S20}$ & $1.040^{+0.091}_{-0.082}$\\
 & $D_{TESS,S21}$ & $1.016^{+0.095}_{-0.088}$\\
 & $D_{TESS,S47}$ & $1.277^{+0.106}_{-0.100}$\\
 & $D_{TESS,S74}$ & $1.011^{+0.093}_{-0.084}$\\
\hline
\sidehead{Planetary Parameters:} 
Mass & $M_p$ (\Mearth / \Mjup) & $112^{+18}_{-19}~/~0.35^{+0.06}_{-0.06}$\\
Radius & $R_p$ (\Rearth / \Rjup) & $9.75^{+0.47}_{-0.47}~/~0.87^{+0.04}_{-0.04}$\\
Density & $\rho_p$ (g/cm\textsuperscript{3}) & $0.661^{+0.160}_{-0.137}$\\
Semimajor Axis & a (au) & $0.0712^{+0.0008}_{-0.0008}$\\
Average Incident Flux\textsuperscript{e} & $\langle F\rangle$ (10\textsuperscript{5} W/m\textsuperscript{2}) & $0.065^{+0.005}_{-0.005}$\\
Planetary Insolation & S (S$_{\oplus}$) & $12.92^{+1.05}_{-1.05}$\\
Equilibrium Temperature & $T_{eq}$ (K) & $528\pm10$\\
\enddata
\tablenotetext{a}{The reported value refer to the 16-50-84\% percentile of the posteriors.}
\tablenotetext{b}{In addition to the absolute RV given in Table~\ref{tab:stellarparam}}
\tablenotetext{c}{Jitter (per observation) added in quadrature to photometric instrument error.}
\tablenotetext{d}{Uncorrected dilution due to the presence of the background stars in the TESS aperture.}
\tablenotetext{e}{We use the solar flux constant 1360.8 W/m\textsuperscript{{-2}} to convert insolation to incident flux.}
\end{deluxetable*}

\section{Discussion}\label{sec:discussion}

\subsection{TOI-5573b in Context with Other GEMS}
\begin{figure}[!htbp]
    \centering
    \includegraphics[width=\linewidth]{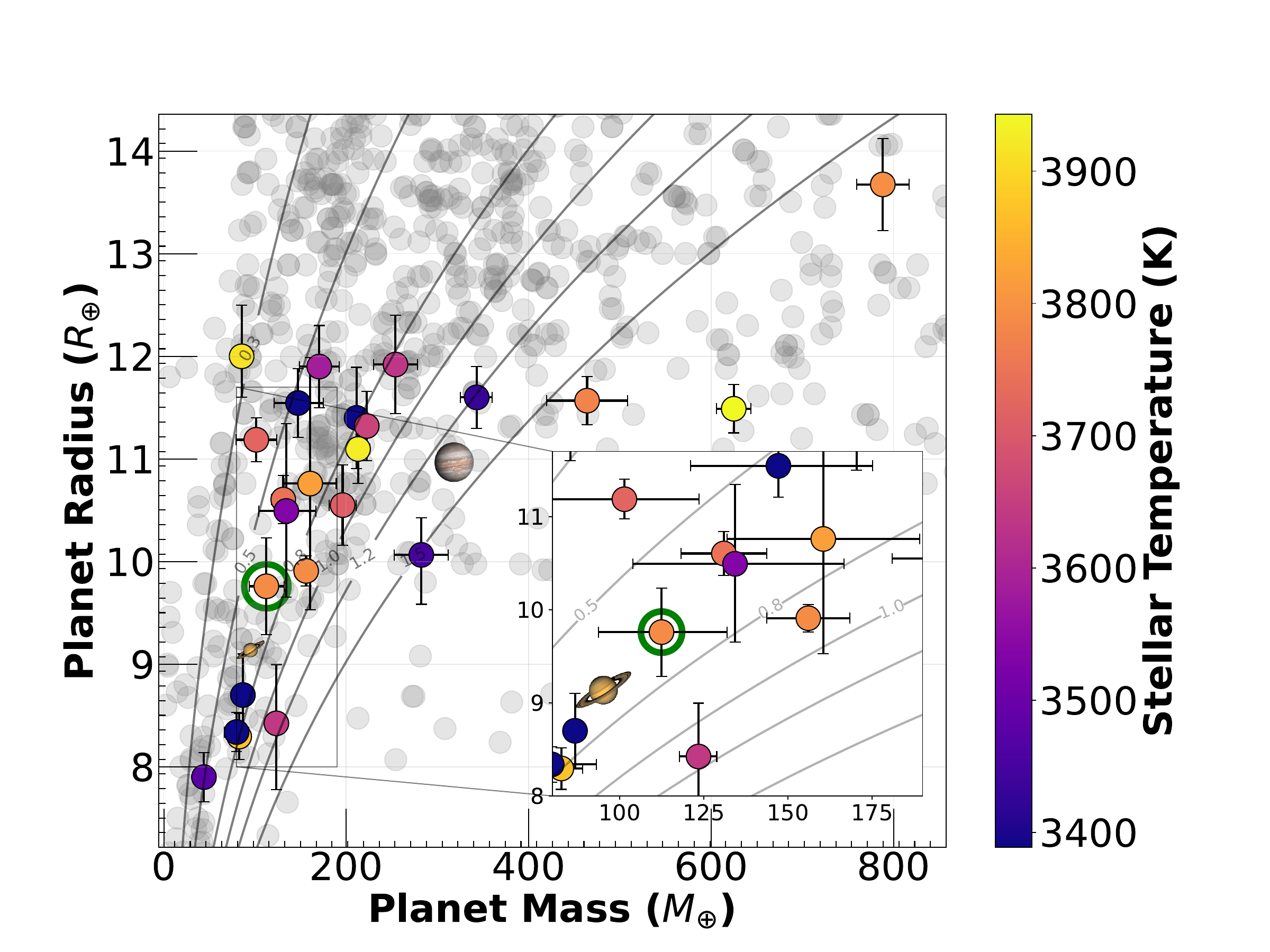}
    \caption{Planet Radius vs. Mass for TOI-5573\,b (highlighted by a green circle) alongside other GEMS with masses between 95–850\,\Mearth. The stellar effective temperature is color-coded for each planet, and for comparison, planets around FGK-type stars are displayed in gray with density contours at 0.3, 0.5, 0.8, 1.0, 1.2, and 1.5 g/cm\textsuperscript{3}}
    \label{fig:mass_Rad_teff}
\end{figure}

\begin{figure}[!htbp]
    \centering
    \includegraphics[width=\linewidth]{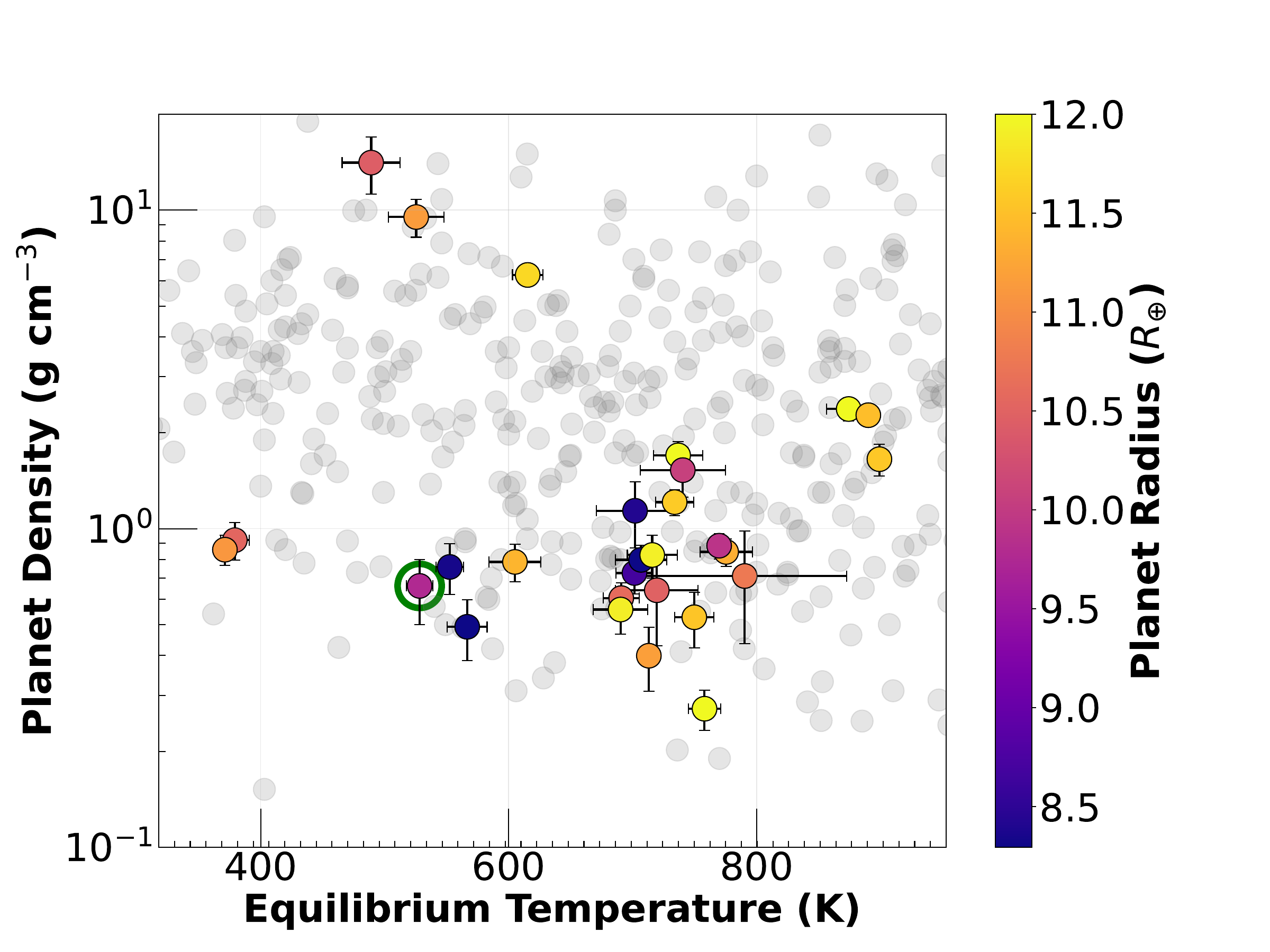}
    \caption{Equilibrium Temperature vs. Planet Density TOI-5573\,b (highlighted by a green circle) alongside other GEMS, with data points color-coded by planet radius, with planets around FGK-type stars are displayed in gray. There are notably no inflated hot Jupiters around M-dwarfs.}
    \label{fig:eqt_dens_rad}
\end{figure}

\begin{figure}[!htbp]
    \centering
    \includegraphics[width=\linewidth]{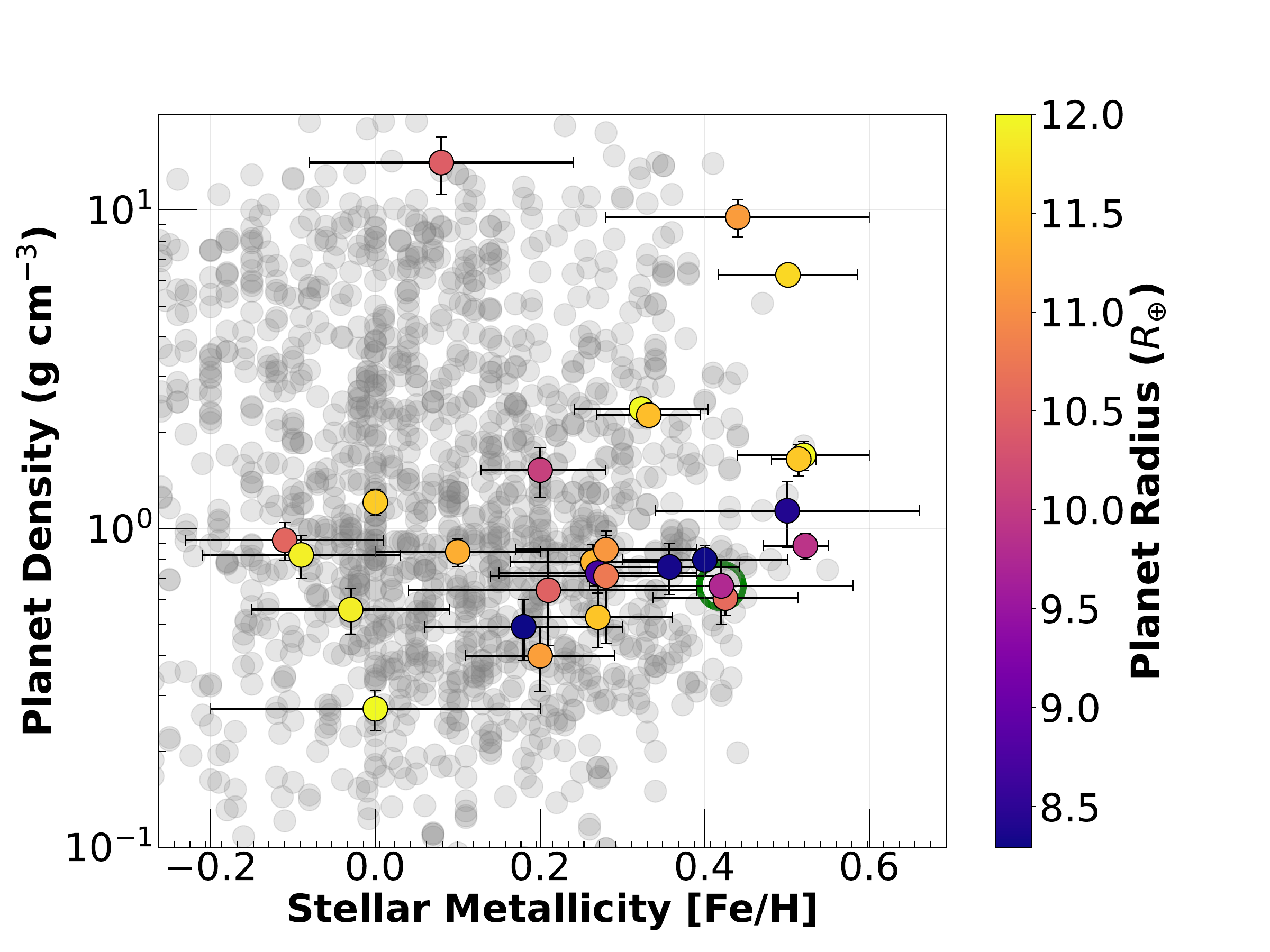}
    \caption{Stellar Metallicity vs. Planet Density TOI-5573\,b (highlighted by a green circle) alongside other GEMS, color-coded by planet radius, with planets around FGK-type stars are displayed in gray}
    \label{fig:met_dens_Rad}
\end{figure}

Around twenty-five transiting GEMS have been identified to date. Here, we explore the position of TOI-5573\,b in the parameter space of transiting GEMS, based on data retrieved from the NASA Exoplanet Archive \citep[NEA;][]{Akeson2013}. The sample is limited to planets with radii between 8 and 15\,\Rearth, and mass measurements with a confidence level above $3\sigma$. Additionally, the host stars of these planets have effective temperatures ranging from 3300 to 4000\,K. As a comparison, transiting giant planets around FGK-type stars are shown in the background of the plots (see Figures~\ref{fig:mass_Rad_teff}, \ref{fig:eqt_dens_rad} and \ref{fig:met_dens_Rad}), allowing for a broader context of planetary characteristics.

TOI-5573\,b has a mass of $112^{+18}_{-19}$\,\Mearth and a radius of $9.75 \pm 0.47$\,\Rearth, making it comparable to Saturn in both size and mass. Planets such as TOI-5344\,b \citep{Han2024}, TOI-5688\,A\,b \citep{Reji2024}, and Kepler-45\,b \citep{Johnson2012} also fall within a $1\sigma$ range of TOI-5573\,b's estimated mass, highlighting the emerging population of Saturn-like planets around M-dwarfs. The density of TOI-5573\,b is calculated to be  $0.661^{+0.160}_{-0.137}$\,g/cm\textsuperscript{3}, which places it within a similar range as Saturn (see Figure~\ref{fig:mass_Rad_teff}), known for its relatively low density compared to Jupiter. The mass-radius relation further solidifies this comparison, as the planets unsurprisingly share not only similar physical characteristics but also comparable densities. Notably, several other planets also exhibit densities between 0.3 and 0.9 g/cm\textsuperscript{3}, emphasizing the prevalence of Saturn-like planets in this region of parameter space.

We also explored the correlation between planetary density and equilibrium temperature (see Figure~\ref{fig:eqt_dens_rad}). The equilibrium temperature is calculated to be $\sim$528\,K, assuming an albedo of zero. We find that TOI-5573\,b is one of the relatively cooler GEMS, sharing its parameter space with planets such as TOI-762\,A\,b \citep{Hartmann2024}, TOI-3984\,A\,b \citep{Canas2023}, HATS-71\,b \citep{bakos2020}, and TOI-3235\,b \citep{Hobson2023}.

We find TOI-5573 to have a super-solar metallicity ($0.42\pm0.16$). The caveats associated with estimating metallicity for M-dwarfs, as discussed in Section~\ref{sec:stellar}, should be kept in mind when interpreting these results. Metallicities were derived either from high-resolution spectroscopy or through SED and photometry, with no distinction made in our plots between these methods.  We find that TOI-5573\,b is one of many transiting GEMS around super-solar metallicity M-dwarfs (see Figure~\ref{fig:met_dens_Rad}). Here, it is important to note that the metallicity is not well constrained, and the typical uncertainties for M dwarfs are large enough that we should be cautious in drawing conclusions. Current approaches, such as spectral template matching, are widely used but not ideal. Several other studies (e.g., \citealt{Han2024}, \citealt{Reji2024}) have used similar methods, and a possible metallicity dependence has also been suggested by \citet{gan2023}. Confirming whether GEMS preferentially form around metal-rich stars will require a line-by-line abundance analysis and a statistically robust comparison with M dwarfs that do not host giant planets. Such an investigation is beyond the scope of this paper.


\subsection{Formation of Saturn-like Exoplanets around M-dwarfs}
Saturn-sized exoplanets orbiting M-dwarfs are typically thought to form through the core accretion model, though their smaller masses compared to Jupiter-sized planets remain an open question. One explanation, proposed by \citet{Movshovitz2010}, suggests that these planets form due to a slowdown in runaway gas accretion, caused by increased opacity in the protoplanetary disk. \cite{Helled2023} refers to these Saturn-like planets as ``failed giants" $-$ they never reached runaway accretion and took several million years to form. In stars with super-solar metallicity, high metallicity can increase disk opacity, reducing the efficiency of heat transfer and slowing gas accretion, preventing the planets from becoming overly massive. Gravitational instability has also been explored as a possible formation pathway, where a massive disk fragments directly into giant planets. Here, we discuss both core accretion and gravitational instability as potential formation mechanisms for TOI-5573\,b.

The core accretion process involves two stages: first, the formation of a solid core by the coagulation of planetesimals and pebbles, typically exceeding 10\,\Mearth, followed by runaway gas accretion onto the core from the protoplanetary disk \citep{pollack1996,Mordasini2008}. For TOI-5573\,b, forming in-situ at its current orbital distance of 8.79\,days seems unlikely due to the high surface density and feeding zone requirements. Instead, it is more plausible that the planet formed farther from the star, beyond 4.5\,au \citep{Reji2024}, where conditions would have allowed core formation, and later migrated inward. The low eccentricity of the planet's orbit suggests migration through interactions with the protoplanetary disk, as gravitational scattering is less likely given that the circularization timescale at TOI-5573\,b's $a/R_* = 25$ far exceeds the age of the universe.


\section{Summary}\label{sec:conclusion}
We present the discovery and confirmation of TOI-5573 b, a Saturn-like exoplanet orbiting an M-dwarf with a period of 8.79\,days. TOI-5573\,b has a mass of $112^{+18}_{-19}$\,\Mearth and a radius of 9.75$\pm$0.47\,\Rearth, resulting in a density of $0.66^{+0.16}_{-0.13}$\,g/cm\textsuperscript{3}. To derive precise planetary and orbital parameters, we modeled both the transit and radial velocity data using Bayesian analysis with Hamiltonian Monte Carlo sampling. This analysis combined 11 transits observed across four TESS sectors (20, 21, 47, and 74), ground-based follow-up from the Red Buttes Observatory, speckle imaging from NESSI, and radial velocity data from HPF and NEID, achieving a 5$\sigma$ precision on the planet's mass.

Saturn-sized exoplanets around M-dwarfs are believed to form via core accretion, though their smaller masses compared to Jupiter-sized planets remain puzzling. Their cores grow large enough to trigger runaway gas accretion and become Jupiters, yet somehow that process is halted. In core accretion theory, dust opacity plays a key role in setting the timescale for hydrogen-helium accretion. High dust content can slow gas accretion, preventing the planets from becoming overly massive \citep{Ikoma2000}. This may explain how Saturn-mass planets form without becoming Jupiters, making them an important population to study separately from both Jupiters and Neptunes, especially if they arise from a finely tuned balance rather than a fundamentally different mechanism \citep{Bodenheimer2025}. Although gravitational instability is another possible formation pathway, it generally leads to larger planets \citep{Hotnisky2024}, making core accretion more plausible for TOI-5573\,b.

TOI-5573\,b has an equilibrium temperature of about 528\,K, making it one of the cooler GEMS. Its temperature and density are comparable to other transiting exoplanets such as TOI-762\,A\,b and TOI-3984\,A\,b. The host star’s super-solar metallicity of $0.42\pm0.16$ suggests that metal-rich environments may play a significant role in shaping planet formation by increasing disk opacity and influencing gas accretion. However, large uncertainties in M-dwarf metallicity estimates, whether derived from high-resolution spectroscopy or SED and photometry, pose challenges. Compared to other GEMS orbiting metal-rich stars, TOI-5573\,b aligns with the observed pattern that giant planets preferentially form around M-dwarfs with super-solar metallicity. Further high-resolution spectroscopic analysis is needed to better understand how stellar metallicity affects the formation and properties of giant exoplanets like TOI-5573\,b, as a more precise determination could provide deeper insights into the role metallicity plays in planet formation.

\software{\texttt{ArviZ} \citep{arviz_2019}, 
AstroImageJ \citep{astroimagej},
\texttt{astroquery} \citep{astroquery},
\texttt{astropy} \citep{astropy1,astropy2,astropy3},
\texttt{barycorrpy} \citep{barrycorrpy},
\texttt{celerite2} \citep{celerite},
\texttt{corner} \citep{corner},
\texttt{EXOFASTv2} \citep{eastman2019exofastv2},
\texttt{exoplanet} \citep{exoplanet_dfm},
\texttt{HxRGproc} \citep{hxrgproc},
\texttt{HPF-SpecMatch} \citep{specmatch},
\texttt{lightkurve} \citep{lightkurve},
\texttt{matplotlib} \citep{matplotlib},
\texttt{numpy} \citep{numpy},
\texttt{pandas} \citep{reback2020pandas},
\texttt{photutils} \citep{larry_bradley_2024_12585239}
\texttt{pyastrotools} \citep{pyastrotools},
\texttt{pyMC3} \citep{salvatier2016probabilistic}, 
\texttt{scipy} \citep{2020SciPy-NMeth},
\texttt{SERVAL} \citep{zechmeister_spectrum_2018},
\texttt{TGLC} \citep{han2023_tglc}
}

\facilities{
Hobby-Eberly Telescope (HET; Habitable-zone Planet Finder (HPF)), WIYN Telescope (WIYN; NN-EXPLORE Exoplanet Investigations with Doppler spectroscopy (NEID), NN-EXPLORE Exoplanet Stellar Speckle Imager (NESSI)), Transiting Exoplanet Survey Satellite (TESS), Red Buttes Observatory (RBO), Zwicky Transient Facility (ZTF), NASA Exoplanet Archive (Exoplanet Archive), Gaia Space Observatory (Gaia), Mikulski Archive for Space Telescopes (MAST)}

\section{Acknowledgments}
C.I.C. acknowledges support from NASA Headquarters through an appointment to the NASA Postdoctoral Program at the Goddard Space Flight Center, administered by ORAU through a contract with NASA.

Z.L.D. would like to thank the generous support of the MIT Presidential Fellowship, the MIT Collamore-Rogers Fellowship, and the MIT Teaching Development Fellowship and to acknowledge that this material is based upon work supported by the National Science Foundation Graduate Research Fellowship under Grant No. 1745302.

These findings stem from observations conducted using the Habitable-zone Planet Finder Spectrograph on the Hobby-Eberly Telescope (HET). We gratefully acknowledge support from various sources, including NSF grants AST-1006676, AST-1126413, AST-1310885, AST-1310875, AST-1910954, AST-1907622, AST-1909506, ATI 2009889, ATI-2009982, AST-2108512, AST-1907622, AST-2108801, AST-2108493 and the NASA Astrobiology Institute (NNA09DA76A), in our efforts to achieve precision radial velocities in the near-infrared (NIR). The HPF team is also appreciative of funding provided by the Heising – Simons Foundation through grant 2017-0494. Furthermore, we acknowledge the collaborative efforts of the University of Texas at Austin, the Pennsylvania State University, Ludwig-Maximilian-Universität München, and Georg-August Universität Gottingen in the Hobby–Eberly Telescope project. The HET is named in recognition of its principal benefactors, William P. Hobby and Robert E. Eberly.
The HET collaboration extends its appreciation to the Texas Advanced Computing Center for its support and resources. We express gratitude to the Resident Astronomers and Telescope Operators at the HET for their adept execution of observations with HPF. Additionally, we acknowledge that the HET is situated on Indigenous land. Furthermore, we wish to recognize and pay our respects to the Carrizo \& Comecrudo, Coahuiltecan, Caddo, Tonkawa, Comanche, Lipan Apache, Alabama-Coushatta, Kickapoo, Tigua Pueblo, and all the American Indian and Indigenous Peoples and communities who have inhabited or become part of these lands and territories in Texas, here on Turtle Island.
We acknowledge the Texas Advanced Computing Center (TACC) at The University of Texas at Austin for providing high-performance computing, visualization, and storage resources that have contributed to the results reported within this paper.

The data presented in this work were acquired at the WIYN Observatory using telescope time allocated to NN-EXPLORE through a scientific partnership involving the National Aeronautics and Space Administration (NASA), the National Science Foundation (NSF), and NOIRLab. Funding support for this research was provided by a NASA WIYN PI Data Award, which is managed by the NASA Exoplanet Science Institute. The observations were conducted with NEID on the WIYN 3.5 m telescope at Kitt Peak National Observatory (KPNO), under NSF's NOIRLab, and were carried out under proposal 2023A-633546 (PI: S. Kanodia), managed by the Association of Universities for Research in Astronomy (AURA) under a cooperative agreement with the NSF. This work was performed for the Jet Propulsion Laboratory, California Institute of Technology, and sponsored by the United States Government under Prime Contract 80NM0018D0004 between Caltech and NASA. The WIYN Observatory is a collaborative effort involving the University of Wisconsin-Madison, Indiana University, NSF's NOIRLab, Pennsylvania State University, Purdue University, University of California-Irvine, and the University of Missouri. We acknowledge the privilege of conducting astronomical research on Iolkam Du’ag (Kitt Peak), a mountain of special significance to the Tohono O’odham people.
Several observations presented in this paper utilized the NN-EXPLORE Exoplanet and Stellar Speckle Imager (NESSI) under the proposal 2022B-936991. NESSI received funding from the NASA Exoplanet Exploration Program and the NASA Ames Research Center. Steve B. Howell, Nic Scott, Elliott P. Horch, and Emmett Quigley built NESSI at the Ames Research Center.

We utilized data from the Gaia mission\footnote{\url{https://www.cosmos.esa.int/gaia}} of European Space Agency (ESA), processed by the Gaia Data Processing and Analysis Consortium (DPAC, \footnote{\url{https://www.cosmos.esa.int/web/gaia/dpac/consortium}}). The DPAC's funding is provided by national institutions, particularly those involved in the Gaia Multilateral Agreement. Additionally, we acknowledge support from NSF grant AST-1907622 for conducting precise photometric observations from the ground.

We express gratitude for the assistance provided by NSF grant AST-1907622 in conducting meticulous photometric observations from the ground.

This study has utilized the Exoplanet Follow-up Observation Program website\footnote{https://exofop.ipac.caltech.edu/tess/}, managed by the California Institute of Technology under contract with the National Aeronautics and Space Administration as part of the Exoplanet Exploration Program.

This work was partially supported by funding from the Center for Exoplanets and Habitable Worlds. The Center for Exoplanets and Habitable Worlds and the Penn State Extraterrestrial Intelligence Center are supported by Penn State and its Eberly College of Science.

\clearpage

\bibliographystyle{apj}
\bibliography{main}

\end{document}